\documentclass[pre,amsmath,preprint,showpacs]{revtex4}
\usepackage{graphicx}
\begin{document}
\title{Steady-state L\'{e}vy flights in a confined domain }
\author{S.~I.~Denisov,$^{1,2,3}$ Werner Horsthemke,$^{4}$
and Peter~H\"{a}nggi$^{1}$} \affiliation{$^1$Institut f\"{u}r Physik,
Universit\"{a}t Augsburg, Universit\"{a}tsstra{\ss}e 1,
D-86135 Augsburg, Germany\\
$^2$Max-Planck-Institut f\"{u}r Physik komplexer Systeme,
N\"{o}thnitzer Strasse~38, D-01187 Dresden, Germany\\
$^3$Sumy State University, 2 Rimsky-Korsakov Street, 40007 Sumy, Ukraine\\
$^4$Department of Chemistry, Southern Methodist University, Dallas, Texas
75275--0314, USA}


\begin{abstract}
We derive the generalized Fokker-Planck equation associated with a Langevin
equation driven by arbitrary additive white noise. We apply our result to study
the distribution of symmetric and asymmetric L\'{e}vy flights in an infinitely
deep potential well. The fractional Fokker-Planck equation for L\'{e}vy flights
is derived and solved analytically in the steady state. It is shown that
L\'{e}vy flights are distributed according to the beta distribution, whose
probability density becomes singular at the boundaries of the well. The origin
of the preferred concentration of flying objects near the boundaries
in nonequilibrium systems is
clarified.
\end{abstract}

\maketitle

\section{INTRODUCTION}

It is remarkable, and at first sight surprising, that a large variety of
physical, biological, financial, and other processes can be described by stable
L\'{e}vy processes with \textit{infinite} variance \cite{1,2,3}. The latter are
defined as continuous-time random processes whose independent and stationary
increments are distributed according to heavy-tailed stable distributions. The
main feature of these distributions is that the tails cannot be cut off or, in
other words, rare but large events cannot be neglected. As a consequence, the
classical stochastic theory, which is based on the ordinary central limit
theorem, is no longer valid.

Due to the heavy-tailed distributions of the increments, stable L\'{e}vy
processes exhibit large jumps, and for this reason these processes are often
called L{\'e}vy flights. L\'{e}vy flights are actually observed in various real
systems. Representative examples include, for instance, fluorescent probes in
living polymers \cite{OBLU}, tracer particles in rotating flows \cite{SWS},
ions in optical lattices \cite{KSW}, cooled atoms in laser fields \cite{BBAC},
subsurface hydrology \cite{8,9}, and ecology \cite{10,11,12}, though a recent
work questions the empirical evidence for L{\'e}vy flights in animal search
patterns \cite{EdPhWaFrMu07}. L\'{e}vy flights have also been predicted for a
large number of model systems \cite{1,2,3}. The ubiquity of these processes is
supported by the generalized central limit theorem \cite{GK}, which states that
all limiting distributions of properly normalized and centered sums of
independent, identically distributed random variables are stable.

The Langevin equation is one of the most important tools for studying noise
phenomena in systems coupled to a fluctuating environment. Introduced by Paul
Langevin just one hundred years ago \cite{Lang} to describe the dynamics of a
Brownian particle, this equation and its various modifications are widely used
in many areas of science \cite{CKW}. The Langevin equation driven by L\'{e}vy
white noise, i.e., noise defined as a time derivative, in the sense of
generalized functions, of a stable L\'{e}vy process, provides a basis for the
study of L\'{e}vy flights in external potentials. It has been shown that the
probability density of L\'{e}vy flights satisfies the fractional Fokker-Planck
(FP) equation \cite{17,18,19,20,21}. The steady-state solutions of this
equation describing \textit{confined} L\'{e}vy flights, i.e., flights with
finite variance, are of particular interest. One reason is that these solutions
will clarify the distribution of flying objects in confined domains. This is an
important issue, especially near impermeable boundaries, in such complex
systems as confined plasmas and turbulent flows. Another reason is that exact
general solutions of a simple form, which are valid for \textit{any} L\'{e}vy
white noise, will be very useful for testing a variety of numerical methods in
this area \cite{22,23,24}. However, known solutions are related to power
potentials and to a very special case of L\'{e}vy white noise with unit index
of stability and zero skewness parameter \cite{25,26}, and thus they are not
suitable for those purposes.

It should be noted that L\'{e}vy white noises do not exhaust all possible white
noises. As a consequence, the fractional FP equation is a particular case of
the \textit{generalized} FP equation, which corresponds to the Langevin
equation driven by an \textit{arbitrary} white noise. Since any white noise is
defined as a time derivative, in the sense of generalized functions, of a
stationary process with independent increments, it can be characterized by the
transition probability density or, alternatively, by the characteristic
function of this white noise generating process. One expects therefore that the
term in the generalized FP equation that describes the effect of the noise on
the dynamics of the system can also be expressed via the characteristic
function. The derivation of the generalized FP equation is of great importance
because it accounts for all possible white noise effects in a unified way and
will be very useful for applications.

In this paper, we put forward the generalized FP equation and find the
analytical solution of this equation for steady-state L\'{e}vy flights in a
confined geometry.

\section{GENERALIZED FOKKER-PLANCK EQUATION}

In many applications, ranging from physical and chemical to biological and
social systems, the relevant degrees of freedom of these systems obey a
(dimensionless) Langevin equation that is equivalent to the equation of motion
for an overdamped particle
\begin{equation}
    \dot{x}(t) = f(x(t),t) + \xi(t).
\label{Langevin}
\end{equation}
Here, $x(t)$ is a particle coordinate, $x(0)=0$, $f(x,t)=-\partial
U(x,t)/\partial x$ is a force field, $U(x,t)$ is an external deterministic
potential, and $\xi(t)$ is a random force (noise) resulting from a fluctuating
environment. Though the quantities in Eq.~(\ref{Langevin}) have different
meanings for different systems, we will use the above terminology to be
concrete.

Under certain conditions (see, e.g., Refs.~\cite{HL,HJ}), the noise $\xi(t)$
can be chosen to be white. In this case, the increment $\delta x(t) = x(t+\tau)
- x(t)$ of the particle coordinate during a time interval $\tau$ ($\tau \to 0$)
is written as
\begin{equation}
    \delta x(t) = f(x(t),t)\tau + \delta\eta(t),
\label{incr}
\end{equation}
which defines the meaning of Eq.~(\ref{Langevin}) in the white-noise
approximation. Here $\delta \eta(t) = \eta(t + \tau) - \eta(t) = \int_{t}^ {t +
\tau} dt'\xi (t')$, and we assume that the integral exists in the mean square
sense. A white noise generating process, i.e., a stationary random process
$\eta(t) = \lim_{\tau \to 0} \sum_{j=0}^ {[t/\tau]-1} \delta \eta(j\tau)$, with
$\eta(0) = 0$ and where $[t/ \tau]$ denotes the integer part of $t/\tau$, is
completely defined by the transition probability density $p(\eta_{j+1},
\tau\vert \eta_{j})$ of a discrete-time process $\eta(n\tau) = \sum_{j=0}^{n-1}
\delta\eta(j\tau)$ ($n \geq 1$) as $\tau \to 0$. Here $\eta_{j+1}$ and
$\eta_{j}$ denote the possible values of $\eta(j\tau + \tau)$ and $\eta(j
\tau)$, respectively. Note that all transition probability densities of the
form $p(\eta_{j+k}, k\tau\vert\eta_{j})$ can be expressed through
$p(\eta_{j+1}, \tau\vert \eta_{j})$ by using the Chapman-Kolmogorov equation
\cite{HL, V-K,HT}. In particular, $p(\eta_{j+2}, 2\tau\vert\eta_{j}) = \int
d\eta_{j+1} p(\eta_{j+2}, \tau\vert \eta_{j+1}) p(\eta_{j+1}, \tau\vert
\eta_{j})$. This implies that the influence of any white noise on the system
can also be fully characterized by the function $p(\eta_{j+1}, \tau\vert
\eta_{j})$. We assume that the transition probability density $p(\eta_{j+1},
\tau\vert \eta_{j})$ is properly normalized, $\int_{-\infty}^ {\infty}
d\eta_{j+1} p(\eta_{j+1}, \tau\vert \eta_{j}) = 1$, and that it satisfies the
condition $\lim_{\tau \to 0} p(\eta_{j+1}, \tau\vert \eta_{j}) =
\delta(\eta_{j+1} - \eta_{j})$, where $\delta(\cdot)$ stands for the Dirac
$\delta$ function. For simplicity, we also assume that $p(\eta_ {j+1}, \tau|
\eta_{j}) = p(\Delta\eta,\tau)$ with $\Delta\eta = \eta_{j+1} - \eta_{j}$.

We define the probability density of the particle coordinate $x(t)$ in the
usual way, namely $P(x,t) = \langle \delta(x - x(t)) \rangle$, where the
angular brackets denote averaging over the noise. Taking the Fourier transform
of $P(x,t)$ according to the definition $\mathcal{F}\{ g(x) \} \equiv g_{k} =
\int_ {-\infty}^{\infty} dx\,e^{-ik x} g(x)$, we obtain $P_{k}(t) = \langle
e^{-ik x(t)} \rangle$, i.e., the characteristic function of $x(t)$. Equation
(\ref{incr}) implies that the increment of this quantity, $\delta P_{k} =
P_{k}(t+\tau) - P_{k} (t)$, can be written in the form $\delta P_{k} =
-ik\tau\langle e^{-ik x(t)} f(x(t),t) \rangle + \langle e^{-ik x(t)}(e^{-ik
\delta \eta(t)} - 1) \rangle$ as $\tau \to 0$. The use of the well-known
properties of the Fourier transform yields $ik \langle e^{-ik x(t)} f(x(t),t)
\rangle = \mathcal{F} \{\partial f(x,t)P(x,t) /\partial x \}$, and the
statistical independence of $x(t)$ and $\delta \eta(t)$ implies \cite{DVH}
$\langle e^{-ik x(t)}(e^{-ik \delta \eta(t)} - 1) \rangle = P_{k}(t) (p_{k}
(\tau) - 1)$, where $p_{k}(\tau) = \mathcal{F}\{ p(\Delta\eta,\tau) \} =
\langle e^{-ik \delta \eta(t)} \rangle$ is the characteristic function of
$\delta\eta(t)$. Dividing $\delta P_{k}$ by $\tau$ and taking the limit $\tau
\to 0$, we obtain the generalized FP equation in Fourier space,
\begin{equation}
    \frac{\partial}{\partial t}P_{k}(t) = -\mathcal{F}\Big\{ \frac{\partial}
    {\partial x}f(x,t)P(x,t)\Big\} + P_{k}(t) \phi_{k},
    \label{F-P0}
\end{equation}
where $\phi_{k} = \lim_{\tau \to 0} (p_{k}(\tau) - 1)/\tau$.

It is advantageous to introduce the characteristic function $S_{k} = \langle
e^{-ik\eta(1)}\rangle$ of $\eta(1)$. We rewrite it as $S_{k} = \lim_{\tau \to
0} (p_{k} (\tau))^ {[1/\tau]}$, using the formula $\eta(1) = \lim_{\tau \to 0}
\sum_{j=0}^ {[1/\tau]-1} \delta \eta(j\tau)$. Replacing $p_{k}(\tau)$ by $1 +
\tau \phi_{k}$ and taking into account that $\lim_ {\varepsilon \to 0} (1 +
\varepsilon)^{1/\varepsilon} = e$, we readily find that $S_{k} = e^{\phi_
{k}}$. Finally, applying the inverse Fourier transform $\mathcal{F}^{-1}\{
g_{k} \} \equiv g(x) = (1/2\pi) \int_{-\infty}^ {\infty} dk\, e^{ik x} g_{k}$
to Eq.~(\ref{F-P0}) and using $\phi_ {k} = \ln S_{k}$, we obtain the
generalized FP equation in real space
\begin{equation}
    \frac{\partial}{\partial t}P(x,t) = -\frac{\partial}{\partial x}
    f(x,t)P(x,t) + \mathcal{F}^{-1}\{P_{k}(t) \ln S_{k}\},
    \label{F-P1}
\end{equation}
with $P(x,0) = \delta(x)$, which corresponds to the Langevin equation
(\ref{Langevin}) driven by an arbitrary white noise.

Equation (\ref{F-P1}) represents our first main result. It constitutes a
closed,  concise representation of the combination of the Fokker-Planck and
Kolmogorov-Feller equations, which are the basic equations governing continuous
and discontinuous Markov processes, respectively \cite{HL,V-K,HT}. A remarkable
feature of this equation is that it accounts for the noise influence in a
\textit{unified} way, namely by means of the characteristic function $S_{k}$ of
the white noise generating process $\eta(t)$ at $t=1$. All presently known FP
equations associated with Eq.~(\ref{Langevin}) can be obtained directly from
Eq.~(\ref{F-P1}). In particular, if $\xi(t)$ is Poisson white noise
characterized by the transition probability density $p(\Delta\eta, \tau) = (1 -
\lambda \tau)\,\delta (\Delta\eta) + \lambda\tau q(\Delta\eta)$, where
$\lambda$ is the average number of jumps of $\eta(t)$ per unit time and
$q(\Delta\eta)$ is the probability density of jump sizes, then $S_{k} =
e^{-\lambda (1 - q_k)}$ and Eq.~(\ref{F-P1}) yields \cite{32,33,34}
\begin{eqnarray}
    \frac{\partial}{\partial t}P(x,t) \!\!&=&\!\!
    -\frac{\partial}{\partial x} f(x,t)P(x,t) - \lambda P(x,t)
    \nonumber\\[6pt]
    &&\!\! + \lambda \int_{-\infty}^{\infty} dy P(y,t)\, q(x-y).
    \label{PoissF-P}
\end{eqnarray}

If $\xi(t)$ is L\'{e}vy white noise then the generalized central limit theorem
\cite{GK} implies that $S_{k}$ is the characteristic function of L\'{e}vy
stable distributions. As is well known (see, e.g., Ref.~\cite{Zol}), the
characteristic function $S_{k}(\alpha, \beta, \gamma, \rho)$ of non-degenerate
stable distributions depends on four parameters: an index of stability $\alpha
\in (0,2]$, a skewness parameter $\beta \in [-1,1]$, a scale parameter $\gamma
\in (0, \infty)$, and a location parameter $\rho \in (-\infty, \infty)$.
Assuming in accordance with the initial condition $P(x,0) = \delta(x)$ that
$\rho =0$ and excluding from consideration the singular case where $\alpha = 1$
and $\beta \neq 0$ simultaneously (in this case $\lvert\phi_{k}\rvert =
\infty$, and the system reaches the final state immediately), we obtain $S_{k}
= S_{k} (\alpha, \beta, \gamma)$, where \cite{Zol}
\begin{equation}
    S_{k}(\alpha, \beta, \gamma) = \exp{\left[ -\gamma \lvert k\rvert^{\alpha}
    \left(1 + i\beta\, \text{sgn}(k) \tan \frac{\pi \alpha}{2}
    \right) \right]}.
    \label{charfunct}
\end{equation}

Equation (\ref{F-P1}) with $S_{k} = S_{k} (\alpha, \beta, \gamma)$ can be
easily rewritten as a fractional differential equation. The Riemann-Liouville
derivatives of a function $g(x)$ on the interval $[-s,s]$ are defined as
\cite{SKM}
\begin{equation}
    _{s}D_{\pm}^{\sigma}g(x) = \frac{(\pm 1)^n}{\Gamma(n - \sigma)}
\frac{d^n}{d x^n} \int_{0}^{s \pm x} dy\, g(x \mp y)\, y^{n - \sigma -1}.
\label{R-L}
\end{equation}
Here, $_{s}D_{+}^ {\sigma}$ and $_{s}D_{-}^{\sigma}$ denote the operators of
the left- and right-hand side derivatives of order $\sigma$ ($0<\sigma <
\infty$), respectively, with $n = 1 + [\sigma]$, and $\Gamma(z)$ is the gamma
function. Since $\mathcal{F} \{_{\infty}D_{\pm}^{ \alpha}P(x,t)\} = (\pm
ik)^{\alpha} P_{k}(t)$, which follows from the definition (\ref{R-L}),
Eq.~(\ref{F-P1}) reduces to the desired fractional FP equation:
\begin{eqnarray}
    \frac{\partial}{\partial t}P(x,t) \!\!&=&\!\! -\frac{\partial}
    {\partial x}f(x,t)P(x,t) - \frac{\gamma}{2\cos(\pi\alpha/2)}
    [(1 + \beta)
    \nonumber\\[6pt]
    &&\!\! \times _{\infty}D_{+}^{\alpha} + (1 - \beta)_{\infty}
    D_{-}^{\alpha}]\,P(x,t).
    \label{F-P2}
\end{eqnarray}
All previously known forms of the fractional FP equation, which correspond to
the Langevin equation (\ref{Langevin}) driven by L\'{e}vy white noise, can be
derived from Eq.~(\ref{F-P2}). In particular, taking into account the relation
$(_{\infty}D_{+}^{ \alpha} +\! _{\infty}D_{-} ^{\alpha})P(x,t) = 2\cos(\pi
\alpha/2)\, \mathcal{F}^{-1} \{\lvert k\rvert^{\alpha} P_{k}(t)\}$ and the
definition of the fractional Riesz derivative \cite{SKM}, $\partial ^{\alpha}
P(x,t) /\partial \lvert x\rvert^{\alpha} = -\mathcal{F}^{-1} \{\lvert
k\rvert^{\alpha} P_{k}(t)\}$, Eq.~(\ref{F-P2}) in the case of symmetric
L\'{e}vy white noise ($\beta = 0$) yields \cite{17,18,19,20,21}
\begin{equation}
    \frac{\partial}{\partial t}P(x,t) = -\frac{\partial}{\partial x}
    f(x,t)P(x,t) + \gamma \frac{\partial^{\,\alpha}}{\partial
    \lvert x\rvert^{\alpha}}P(x,t).
    \label{F-P3}
\end{equation}
Specifically, if $\xi(t)$ represents Gaussian white noise of intensity $D$,
i.e., $\alpha = 2$ and $\gamma = D$, then Eq.~(\ref{F-P3}) becomes the ordinary
FP equation \cite{Risk}.

We note that Eq.~(\ref{F-P1}) also applies to the case of compound white noises.
For example, if $\xi(t)$ is the sum of independent L\'{e}vy and Poisson white
noises then $S_{k} = S_{k} (\alpha, \beta, \gamma)\, e^{-\lambda (1 - q_k)}$
and Eq.~(\ref{F-P1}) can be written as
\begin{eqnarray}
    \frac{\partial}{\partial t}P(x,t) \!\!&=&\!\! -\frac{\partial}
    {\partial x}f(x,t)P(x,t) - \frac{\gamma}{2\cos(\pi\alpha/2)}
    [(1 + \beta)
    \nonumber\\[6pt]
    &&\!\! \times _{\infty}D_{+}^{\alpha} + (1 - \beta)_{\infty}
    D_{-}^{\alpha}]\,P(x,t)
    \nonumber\\[6pt]
    &&\!\! - \lambda P(x,t) + \lambda \int_{-\infty}^{\infty}\! dy
    P(y,t)\, q(x-y).\quad \;
    \label{F-P4}
\end{eqnarray}

\section{STEADY-STATE L\'{E}VY FLIGHTS IN A CONFINED GEOMETRY}

We apply Eq.~(\ref{F-P2}) to the case of stationary L\'{e}vy flights in an
infinitely deep potential well. We assume that $f(x,t)=0$ within the well,
i.e., for $x \in [-l,l]$, and that the boundaries at $x = \pm l$ are
impermeable for particles, i.e., $P(x,t) = 0$ at $\lvert x\rvert > l$. With
these conditions, Eq.~(\ref{F-P2}) for the stationary probability density
$P_{\text{st}}(x)$ reduces to $(1 +\beta)_{l}D_{+}^{\alpha} P_{\text{st}} (x) +
(1 - \beta)_{l} D_{-}^{\alpha} P_{\text{st}}(x) = 0$. Rewriting this equation
as $dJ(x)/dx = 0$, where $J(x)$ is the probability current, and using the
boundary condition $J(\pm l)=0$ \cite{Risk}, we obtain the equation $J(x)=0$,
which for $0 <\alpha <1$ reads
\begin{equation}
    (1+\beta)\!\int_{-l}^{x}dy\frac{P_{\text{st}}(y)}{(x-y)^{\alpha}} -
    (1-\beta)\!\int_{x}^{l}dy\frac{P_{\text{st}}(y)}{(y-x)^{\alpha}} = 0.
    \label{F-Pst2}
\end{equation}

The fact that $_{l}D_{\pm}^{\alpha} (l \pm x)^{\alpha - 1} \equiv 0$ \cite{SKM}
suggests seeking a solution of Eq.~(\ref{F-Pst2}) in the form $P_{\text{st}}
(x) = C(l+x)^ {-\nu} (l-x) ^{-\mu}$, where $C$ is a normalization factor. The
parameters $\nu$ and $\mu$ are determined by the equation
\begin{eqnarray}
    &(1\!+\!\beta)\text{B}(1\! - \!\alpha,1\! - \!\nu)z^{1 - \alpha}
    F(1\! - \!\alpha, \mu;2\! - \!\alpha\! - \!\nu;-z)\quad\quad&
    \nonumber\\[4pt]
    &\;=(1\! - \!\beta) \text{B}(1\! - \!\alpha,1\! - \!\mu)
    F(1\!-\!\alpha,\nu;2\!-\!\alpha\!-\!\mu;-z^{-1}).\quad\;&
    \label{rel1}
\end{eqnarray}
Here, $z=(l+x)/(l-x)$, $\text{B}(a,b) = \Gamma(a)\Gamma(b)/\Gamma(a+b)$ is the
beta function, and $F(a,b;c;y)$ is the Gauss hypergeometric function. Equation
(\ref{rel1}) must be independent of $x$, since $\nu$ and $\mu$ do not depend on
$x$. This requirement leads to the condition $\alpha + \nu + \mu=2$. Using the
relation $F(a,b;b;-y) = (1+y)^{-a}$ \cite{BE}, we find that in this case
Eq.~(\ref{rel1}) becomes
\begin{equation}
    (1+\beta)\text{B}(1-\alpha,1-\nu) - (1-\beta)\text{B}(1-\alpha,1-\mu) = 0.
    \label{rel2}
\end{equation}
Solving the equations $\alpha + \nu + \mu=2$ and (\ref{rel2}) with respect to
$\mu$ and $\nu$, we find
\begin{equation}
    \bigg\{ \begin{array}{ll}
    \mu \\
    \nu
    \end{array} \bigg\}=
    1 - \frac{\alpha}{2} \pm \frac{1}{\pi} \arctan\left(\beta \tan\frac{\pi
    \alpha}{2}\right),
    \label{nu_mu}
\end{equation}
where $\arctan x$ denotes the principal value of the inverse tangent function.
Similar calculations for $1 < \alpha \leq 2$ lead to the same result, and
formula (\ref{nu_mu}) is valid for all $\alpha \in (0,2]$ (excluding the case
$\alpha = 1$, $\beta \neq 0$). Finally, calculating the normalization factor
$C$, we obtain
\begin{equation}
    P_{\text{st}}(x) = (2l)^{1-\alpha} \frac{(l+x)^{-\nu} (l-x)^{-\mu}}
    {B(1-\nu,1-\mu)}.
    \label{Pst}
\end{equation}

Equations (\ref{nu_mu}) and (\ref{Pst}) represent our second main result.
Particles that perform L\'{e}vy flights are distributed in an infinitely deep
well according to the beta distribution (see Fig.~1). The main feature of this
distribution is the singular behavior of $P_{\text{st}} (x)$ as $\lvert x\rvert
\to l$ if $\alpha <2$ and $\beta \neq \pm 1$. The reason is that for $\alpha
<2$ the particles can perform random jumps in both directions. However, the
boundaries are impermeable, and consequently the particles are concentrated
preferably near these two boundaries. In particular, for $\beta = 0$,
Eqs.~(\ref{nu_mu}) and (\ref{Pst}) yield $P_{\text{st}} (x) = (2l)^{1 - \alpha}
\Gamma (\alpha) (l^{2} - x^{2})^{\alpha/2 - 1}/ \Gamma^{2} (\alpha/2)$. In
contrast, for $\beta = \pm 1$, one-sided jumps dominate and the particles are
concentrated near one of the boundaries. Specifically, $P_{\text{st}}(x) =
\delta(l - \text{sgn} (\beta)x)$, if $0 <\alpha < 1$, and $P_{\text{st}}(x) =
(2l)^{1 - \alpha} (\alpha - 1) (l + \text{sgn} (\beta) x)^{\alpha - 2}$, if $1
<\alpha <2$. Finally, for $\alpha = 2$ the sample paths of $x(t)$ are
continuous and the stationary distribution is uniform, i.e., $P_{\text{st}}(x)
= 1/(2l)$.

\section{CONCLUSIONS}

We have derived the generalized FP equation associated with the Langevin
equation driven by an \textit{arbitrary} white noise. This FP equation accounts
for the influence of the noise by means of the characteristic function of the
white noise generating process. In the case of L\'{e}vy flights, this equation
has been reduced to a fractional FP equation and has been solved analytically
in the steady state for a confined domain. It has been shown that both
symmetric and asymmetric L\'{e}vy flights in an infinitely deep potential well
are distributed according to the beta probability density. The preferred
concentration of flying objects near impenetrable boundaries results from the
jumping character of L\'{e}vy flights.

\section*{ACKNOWLEDGMENTS}

The authors are grateful to A. Dubkov and I. Goychuk for useful discussions.
S.I.D. acknowledges the support of the EU through Contract No.
MIF1-CT-2006-021533 and P.H. acknowledges financial support by the Deutsche
Forschungsgemeinschaft via the Collaborative Research Centre SFB-486, Project
No. A10, and by the German Excellence Cluster {\it Nanosystems Initiative
Munich} (NIM).


\begin{figure}[ht]
    \centering
    \includegraphics[totalheight=6.5cm]{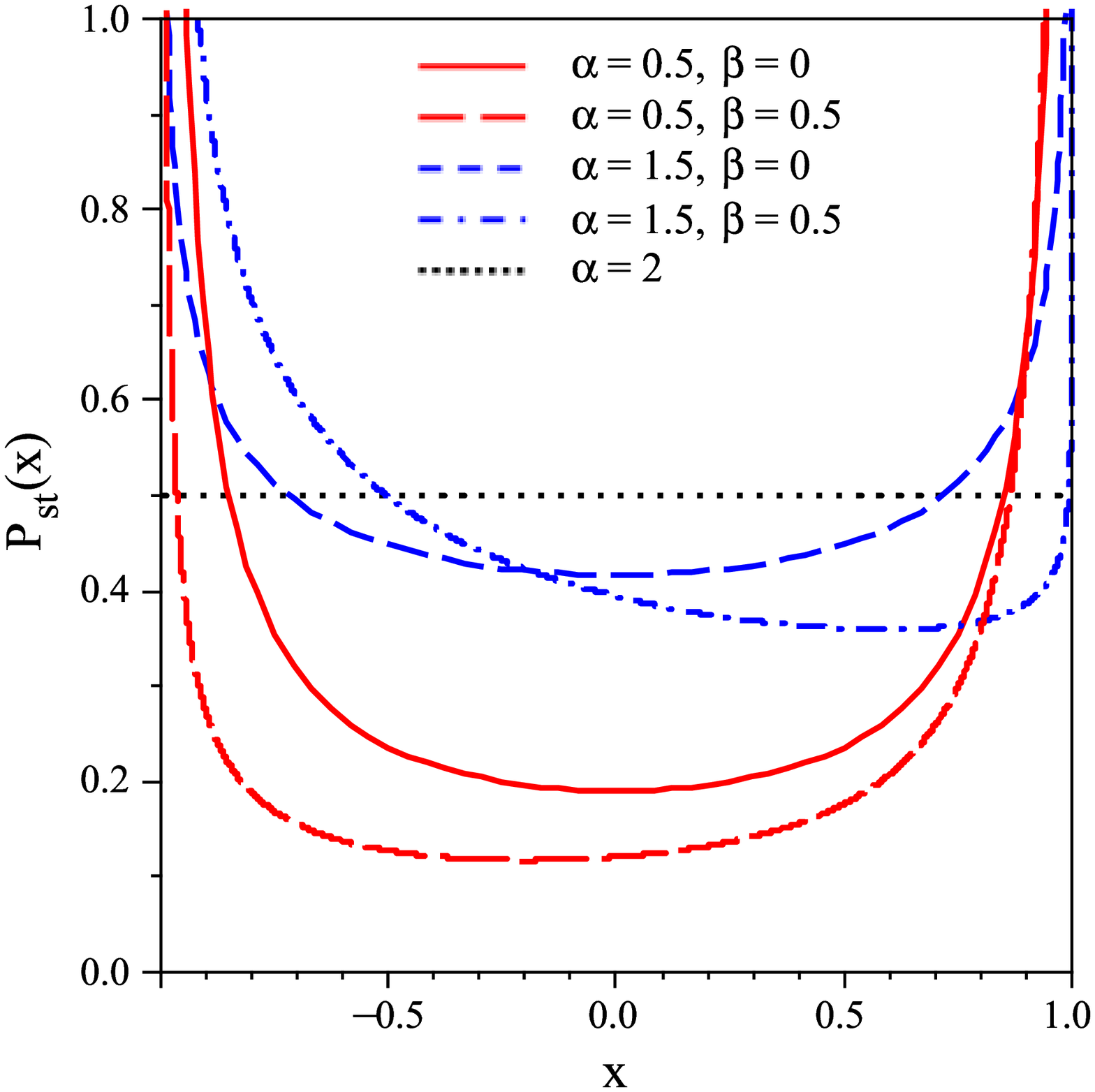}
    \caption{\label{fig1} (Color online) Plots of the stationary
    probability density (\ref{Pst}) for different values of the
    parameters $\alpha$ and $\beta$ and $l = 1$. }
\end{figure}

\end{document}